\documentclass[12pt]{article}
\usepackage{amssymb}
\usepackage{bigints}
\usepackage{amsmath}
\usepackage{amsthm}
\usepackage{empheq}
\usepackage{framed}
\usepackage{a4wide}
\usepackage{enumerate}
\usepackage{makeidx}
\usepackage{graphicx}
\usepackage{caption}
\usepackage{subcaption}
\usepackage{ifpdf}
\usepackage{setspace}
\makeindex
\usepackage{bm}
\usepackage{epsfig}
\usepackage{graphics}
\usepackage{colortbl}
\usepackage{colordvi}
\usepackage{verbatim}
\usepackage{amsmath}
\usepackage{multirow} 
\usepackage{booktabs} 
\newcommand{\be}{\begin{equation}}
\newcommand{\ee}{\end{equation}}
\newcommand{\ba}{\[\begin{aligned}}
\newcommand{\ea}{\end{aligned}\]}
\newcommand{\bea}{\begin{eqnarray}}
\newcommand{\eea}{\end{eqnarray}}
\newcommand{\beann}{\begin{eqnarray*}}
\newcommand{\eeann}{\end{eqnarray*}}
\newcommand{\bs}{\begin{split}}
\newcommand{\es}{\end{split}}
\theoremstyle{definition}

\newcommand*{\cD}{\mathcal{D}}
\newcommand*{\cE}{\mathcal{E}}

\newcommand*{\cH}{\mathcal{H}}

\newcommand*{\vpl}{v_{||}}
\newcommand*{\vpls}{v_{||*}}
\newcommand*{\dpl}{\nabla_{||}}
\newcommand*{\dl}{\bm{\nabla}}

\newcommand*{\ExB}{\bm{E}\times \bm{B}}
\newcommand*{\Fpr}{F_\perp}
\newcommand*{\Tpr}{\tau_\perp}

\newcommand*{\zth}{\bm{\nabla} \zeta}
\newcommand*{\grpsi}{\bm{\nabla} \psi}

\newcommand*{\ep}{\epsilon}

\newcommand*{\dt}{\mathrm{d}}
\newcommand*{\Iph}{I \varphi'}
\newcommand*{\IphB}{\frac{I \varphi'}{B}}

\newcommand*{\cEs}{\mathcal{E}_{*}}
\newcommand*\at[2]{\left.#1\right|_{#2}}

\let\oldhat\hat
\renewcommand{\vec}[1]{\mathbf{#1}}
\renewcommand{\hat}[1]{\oldhat{\mathbf{#1}}}

\newcommand*{\assign}{\ensuremath{\kern.5ex\raisebox{.1ex}{\mbox{\rm:}}\kern -.3em =}}
\usepackage{graphicx}
\makeatletter
\newcommand{\shorteq}{%
  \settowidth{\@tempdima}{-}
  \resizebox{\@tempdima}{\height}{=}%
}
\title{Effective mass in Rosenbluth-Hinton type zonal flows }
\author{W Sengupta, A.B. Hassam \footnote{IREAP, University of Maryland,College Park}}
\date{\vspace{-5ex}}
\begin{document}
\maketitle
\begin{abstract}
 
An initial radial electric field, $E_r(0)$, in an axisymmetric tokamak, results in geodesic acoustic mode (GAM) oscillations. The GAMs Landau damp, resulting in a much smaller final residual electric field, $E_r(\infty)$, and accompanying parallel zonal flows (Rosenbluth and Hinton, 1998 PRL 80, 724, hereafter RH). The phenomenon exhibits a large effective mass (inertia due to flows), with an enhancement of order the well-known RH factor.  In apparent paradox, the final angular momentum in the RH parallel zonal flow is much smaller than the angular momentum expected from the well-known rapid precession of the trapped particle population in the final electric field.  In addition, an effective mass calculated naively based on the rapid trapped particle (TP) precession is much larger than the RH factor.  A drift kinetic calculation is presented showing that the mathematical origin of the extra mass factor is a shift, proportional to $E_r$, of the usual energy coordinates in phase space.  Importantly, this shift contributes to the effective mass even if the system is linearized in $E_r$, and can be interpreted as a first order linear shift in the Jacobian. Further, the Jacobian shift recovers the large TP precession flow and also uncovers the presence of reverse circulating particle flows that, to lowest order, are equal and opposite to the TP precession. A detailed calculation is presented.  
\newline Work supported by DOE.
\end{abstract}

\section{Introduction and Motivation}
\label{sec:intro}
\begin{figure}[h]
\begin{center}
\includegraphics[height=0.5\textwidth,width=0.6\textwidth,angle=0]{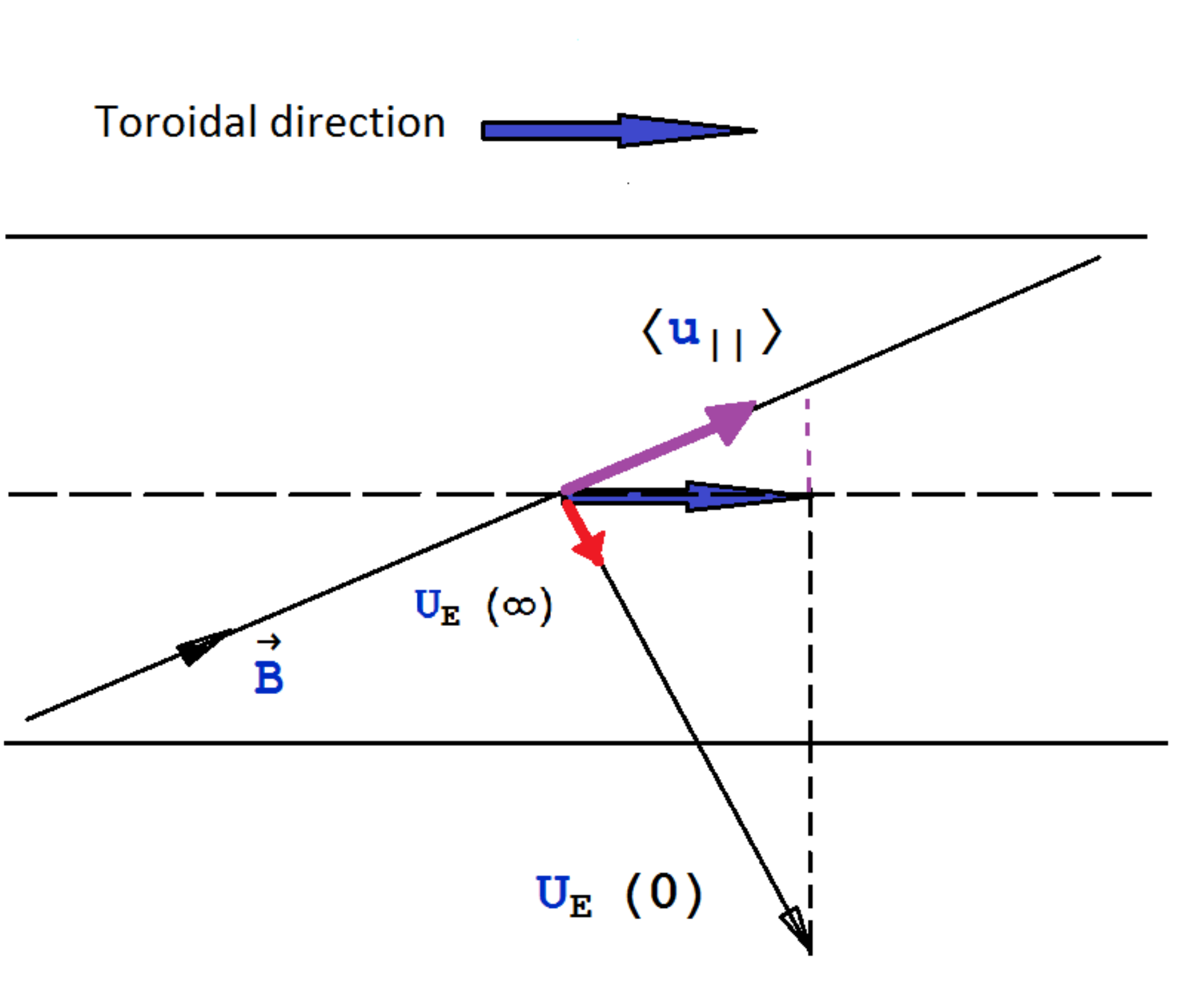}
\caption{Comparison of initial and final RH flows}
\label{fig:RHflows}
\end{center}
\end{figure}

An initial radial electric field, $E_r(0)$, in an axisymmetric tokamak results in GAM oscillations. In a collisionless system, the GAMs Landau damp. However, it was shown by Rosenbluth and Hinton (RH)\cite{RH} that, in the asymptotic steady state, there persists a residual electric field, $E_r(\infty)$, and an associated parallel zonal flow. RH showed that the initial and final electric fields are related according to $E_r(0)= (1+\cD)E_r(\infty)$, where $\cD \sim 1.6 q^2/\sqrt{\ep}$.  Here, $q$ is the tokamak safety factor and $\ep \ll 1$ is the tokamak inverse aspect ratio.  Since the initial $\ExB$ flow, $U_E(0)$, has a toroidal component and, thus, an initial toroidal angular momentum, and the final $\ExB$ flow is much smaller than the initial, a substantial parallel zonal flow must arise in order to preserve angular momentum (see Figure \ref{fig:RHflows}).  The size of the  parallel zonal flow, as found by RH, can be deduced from the geometry of Figure \ref{fig:RHflows} to be of order $(\epsilon/q)U_E(0)$. Finally, as we will elaborate later, the term $1+\cD$ is like an effective mass, arising from the inertia due to the parallel flows.

 A question arises when one considers the individual contributions to the angular momentum of the trapped and circulating fractions of the plasma. It is well known that in the presence of a radial electric field, trapped particles precess toroidally. The speed of precession is of order $(q/\epsilon)U_E$ and represents a rapid rate inasmuch as it is much larger than $U_E $.  In apparent paradox, one finds that the angular momentum in the TP population precessing in the final RH electric field is much larger than the total final RH angular momentum (the TP precession angular momentum is of order ${\sqrt{\ep}}(q/\epsilon)U_E(\infty)$, while the RH calculated final angular momentum is of order $(\epsilon/q)U_E(0)$, as discussed above.  Here we have accounted for the lower density of the trapped fraction, i.e, $n^{TP} \sim O(\sqrt{\epsilon})$.)  In addition, it is reasonable to expect (as we describe below) that such a large precession kinetic energy could result in a fractional effective mass factor of order $\sim \left(\frac{1}{2} (nU^2_{||})^{TP}/\frac{1}{2}n U_E^2\right) \sim O(n^{TP} q^2/\ep^2)$. This factor is, in fact, larger than the RH mass factor, by $1/\epsilon$.
 
\begin{figure}[h]
\begin{center}
\includegraphics[height=0.4\textwidth,width=0.4\textwidth,angle=0]{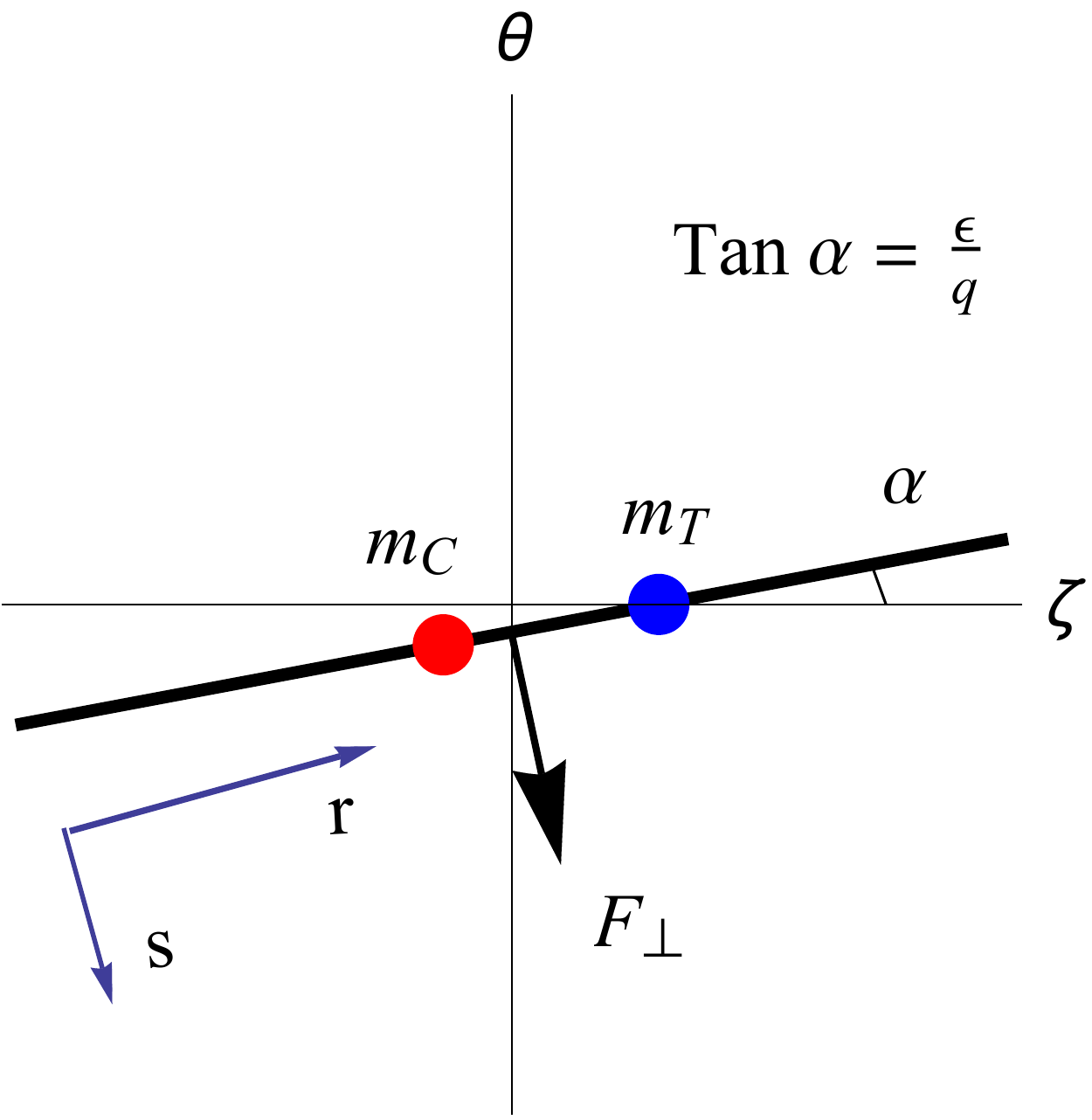}
\caption{A toy model}
\label{fig:toy}
\end{center}
\end{figure}
A simple toy model can be constructed to illustrate these points. We consider a massless rod and two beads of masses $m_T,m_C$ that can slide freely, without interaction, along the rod; one of them ($m_T$) is further constrained in that it can only move horizontally, that is to say it stays trapped inside a linear horizontal 1D channel. This system is depicted in Fig \ref{fig:toy}. The rod represents a magnetic field line; $m_C$ represents circulating particles (CPs), while $m_T$ represents deeply trapped particles. The rod is inclined at a small angle given by $\sin{\alpha} = \epsilon/q \ll 1$. Consider now an external perpendicular force, $F_\perp$, acting on the rod, as shown in the figure.  We want to obtain the effective mass of the system defined according to the constrained Newton's equation $M\ddot{s}= F_\perp$, where $s$ is the distance measured along $F_\perp$ and $\dot{s}$ is the speed of the rod in the lab frame. The Lagrangian for this system is
 \begin{align*}
  L(s,\dot{s},\dot{r})= \frac{1}{2 \sin^{2}\alpha}\:(m_T+m_C)\:\dot{s}^2 +\frac{1}{2}m_C \:( \dot{r}^2+ 2 \dot{s} \: \dot{r} \cot{\alpha})+F_\perp s
  \end{align*}
leading to the equation of motion,
\begin{align*}
\ddot{s} = \frac{F_\perp}{m_C+m_T/ \sin^2{\alpha}} \quad \Rightarrow \quad M= m_C+m_T/ \sin^2{\alpha} \quad = m_C+m_T \frac{q^2}{\epsilon^2}.
\end{align*}
This shows that the effective mass from the constrained mass $m_T$ is $m_T (q^2/\ep^2)$, illustrating our conjecture for TPs above.
 
This line of investigation raises further questions when one calculates separately the CP and TP flows associated with the residual RH zonal flows: as we will show, by direct calculation \cite{Xiao} based on standard drift-kinetic theory, we find the RH parallel flow for the TP's to be of $O(q\: U_E)$, smaller than the precession drift by $1/\ep$. In addition, a direct calculation of the net flux surface averaged poloidal flow of the TP's surprisingly gives a nonzero result, namely, a net poloidal flow of $O(q\: U_E)$.
 
 We note that the RH problem, as posed by Rosenbluth and Hinton, starts with an initial electric field that sets up GAMs, eventually settling to a steady residual flow. Our toy problem, as posed, does not incorporate GAMs, and is based on an external driver force $F_\perp$. Nonetheless, as we will show later, the externally driven problem is a relevant comparison.  In particular, one may revisit the RH problem as the tokamak plasma response to a weak external perpendicular force; in that case, we will show that the same RH factor, or effective mass, is obtained. Such a force could arise from perpendicular neutral beams, for example:  the force would provide toroidal torque that would slowly increase the angular momentum. There are accompanying GAMs, but of negligible amplitude. The final electric field is once again reduced by the same factor $M=1+\cD$.
 
 Our study in this paper is motivated by an attempt to understand the discrepancy in the flows as well as in the naively expected effective mass of the RH problem.   The discussion below is organized as follows: In section \ref{sec:KE}, we present the basic system of equations, consisting of the drift kinetic equation and the angular momentum conservation equation in axisymmetric toroidal geometry. We then solve the classic Rosenbluth-Hinton problem in section \ref{sec:RHproblem} and point out the aforementioned discrepancies in the flows. We introduce the shifted coordinates in section \ref{shift} and redo the RH problem in these new coordinates in section \ref{sec:RHsproblem}, to reconcile trapped particle toroidal precession in RH flows. In section \ref{sec:barelyCP}, we illustrate the role of barely circulating particles in cancelling the large trapped particle precession and thereby explain the smaller overall RH effective mass. We summarize our results in section \ref{sec:summary} and discuss future lines of research. 
 
\section{Kinetic Equations}
\label{sec:KE}
We begin our calculation with the drift kinetic equation (DKE) as formulated by Kulsrud, Frieman, and Hinton-Wong \cite{Kulsrud,Frieman,HW}. The DKE is derived in ``MHD ordering" and thus allows large, sonic level $\bm{E}\times \bm{B}$ flows. A consistent ordering also requires that the parallel electric field $E_{\parallel}$ be very small compared to $E_{\perp}$.   In the electrostatic limit ($\partial_t \ll V_A/L_{||}$) the full DKE is given by
\be
\frac{\partial f}{\partial t}+ (\vec{U_E} + v_{||}\hat{b})\cdot \bm{\nabla} f + 
 \left( -\hat{b}\cdot \left( (\vec{U_E} + v_{||}\hat{b})\cdot \bm{\nabla} \right) \vec{U_E} + \mu B \:\bm{\nabla }\cdot  \hat{b}+\frac{e}{m}E_{\|} \right)\frac{\partial f}{\partial v_{||}} =0
\label{full DKE} 
 \ee
\noindent where,
\be 
\vec{U_E}=\frac{\vec{B}\times \bm{\nabla} \psi}{B^2}\varphi'(\psi)
\label{UE}
\ee
\noindent is the $\bm{E} \times \bm{B}$ flow and $f=f(v_{||},\mu, \vec{x},t)$. The magnetic field $\bm{B}$ is defined as usual by $\bm{B} = I \zth + \zth  \times \grpsi$.   Here, the $E_{\parallel}$ force term is of the same order as the other parallel force terms (the mirror force and inertial forces) in the equation. The above DKE applies for both ions and electrons, though we will assume small electron mass and thus the electron response will be taken to be adiabatic.  The full system in the electrostatic approximation consists of four variables, namely, the two distribution functions, the potential $\varphi$, and $E_{\parallel}$.   These four unknowns are governed by the two DKEs, the quasineutrality condition $n_e = n_i$, and the equation of conservation of angular momentum, namely \cite{cglgam}
\begin{align}
\partial_t\left\langle n \varphi' \frac{|\bm{\nabla}\psi|^2}{B^2}   -  \int  d^3v  \frac{I \vpl}{B} f\right\rangle= \tau_\perp   \label{AngMom1}
\end{align}
\noindent where $\left\langle \cdot \right\rangle$ represents a flux surface average, and $\tau_\perp = \left\langle n  \frac{|\bm{\nabla}\psi|}{B} \frac{F_\perp}{m}\right\rangle$ is a toroidal torque due to a perpendicular force $F_\perp$. The latter represents an external force, such as from a neutral beam, that could accelerate the $\bm{E}\times \bm{B}$ flow.  It can be shown that in axisymmetric geometry, the equation governing the angular momentum is identical to the radial current quasineutrality condition. This equivalence is shown in Appendix \ref{App:AppendixB}.   For the present purposes, we will find the former equation to be more convenient.

In this paper, we will only be concerned with time scales which are subsonic, i.e., $d/dt << c_s/qR$.   In this limit, as we will show more precisely later, $E_{\parallel}$ is small and can be neglected.   In that case, the system can be closed by simply using the DKE for ions, Eq. (\ref{full DKE}), and the angular momentum conservation equation (\ref{AngMom1}).  As a further simplification, we will order $q \gg 1 $ but $U_E \sim v_{th}/q$.   In this ordering, the nonlinear in $U_E$ terms in the DKE can be neglected compared with cross terms in $\vpl$ and $U_E$, since $ |\vpl \vec{b}\vec{b} : \nabla U_E| : |b U_E : \nabla U_E|   \sim  1: \frac{1}{q}$. Given these orderings, Eq.(\ref{full DKE}) can be recast as
\begin{equation}
\frac{\partial f}{\partial t}+ (\vec{U_E} + v_{||}\hat{b})\cdot \bm{\nabla} f + 
 \left(v_{||} \vec{U_E}\cdot \bm{\kappa} - \mu  \bm{\dpl }B \right)\frac{\partial f}{\partial v_{||}} =0.
 \label{DKE0}                                                           
\end{equation}
\noindent  We now use the form for $\vec{U_E}$, Eq. (\ref{UE}), to simplify (\ref{full DKE}).   In particular, $\vec{U_E} \cdot \kappa  = \vec{U_E} \cdot \nabla B/B$, in the low $\beta$ limit.  We also assume axisymmetry and thus $\bm{B} \times \grpsi \cdot \dl = I B \dpl$. Given these, (2) can be recast to the form 
 
\begin{align}
\frac{\partial f}{\partial t}+ \left(\vpl +\frac{\Iph}{B}\right)\dpl f + 
 \left(v_{||}\frac{\Iph}{B} - \mu B \right)\frac{\dpl B}{B}\frac{\partial f}{\partial v_{||}} =0.
 \label{axiDKE}    
\end{align}
\noindent where $f=f(\vpl,\mu, \bm{x},t)$.  We will now use (\ref{axiDKE}) and (\ref{AngMom1}) as the closed set of equations for the two variables $f$ and $\varphi$.

\section{The classic Rosenbluth Hinton problem}
\label{sec:RHproblem} 
We begin by reviewing the RH problem.   We are interested mainly in the effective mass physics as derived by RH.  This physics can be recovered by setting the field lines into motion by applying a weak perpendicular external force $F_{\perp}$.  Since the force is weak, we look for a sub-bounce frequency solution according to the ordering $\partial /\partial t \sim \vec{U_E} \cdot \nabla \sim F_{\perp} << \vpl  \dpl$. To lowest order from (\ref{axiDKE}), we have    
\begin{align}
\vpl\dpl f_0 - \mu \frac{\dpl B}{B}\frac{\partial f_0}{\partial v_{||}} =0    
\end{align}
\noindent which yields $f_0=f_0(\cE,t)$, where $\cE = \vpl^2 +\mu B$ is the energy.    The lowest order angular momentum equation from (\ref{AngMom1}) is simply the 2nd term on the LHS set to zero.   This is identically satisfied if we assume that $f_0(\cE,t)$ is symmetric in $\vpl$ with respect to the circulating particles.   To first order, Eq (\ref{axiDKE}) becomes
\begin{align}
\frac{\partial f_0}{\partial t}+ \vpl \dpl' f_1 = 
 v_{||}\dpl'\left(\frac{\Iph}{B}\vpl \right)\frac{\partial f_0}{\partial \cE}
\end{align}
 \noindent where we have transformed from $\vpl$ to $\cE$ coordinates with ${\dpl}^{\prime}$ being the gradient operator at constant $\cE$, and we have used $\vpl \dpl^\prime (\vpl) = -\mu \dpl B $.   Annihilating the $f_1$ term by bounce averaging as in $\overline{f} = \oint (dl f /\vpl)/ \oint (dl/\vpl)$, we get $f_0=f_0(\cE)$ and
 
\begin{align}
f_1=I{\varphi}'\left( \frac{\vpl}{B} - g \right) {f_0}^\prime
\end{align}
\noindent where $g(\cE)$ is yet to be determined. To second order we have,
\begin{align}
\frac{\partial f_1}{\partial t}+ \vpl \dpl f_2 = 0. 
\label{eqnf2}
\end{align}
\noindent Annihilation of Eqn(\ref{eqnf2}) yields $\overline{\partial_t {f}_1}=0$, which gives $g=\overline{\left( {\vpl}/{B}\right)} $.  Thus, we get the RH solution for $f_1$ correct to first order\cite{Xiao}, viz.
\begin{align}
f_1=\IphB\left(
\frac{\vpl}{B}-\overline{\left(\frac{\vpl}{B}\right)}\right) {f_0}^{\prime} .
\end{align}
 
\noindent   We would now insert $f_1$ into the second term of the angular momentum equation, Eq (\ref{AngMom1}).  We would thus need to calculate the parallel flow to first order, viz.,  
\be
n_0 u_{||1}= \int d^3v\: v_{||} f_0 =  I \varphi' \int \,d^3v\,\,   \vpl \left( \frac{v_{||}}{B}-\overline{\left(\frac{v_{||}}{B}\right)}\right) \frac{\partial f_0}{\partial\cE}
\label{nupl}
\ee

\noindent  Using an expansion in $\epsilon$, we find 
\be
{u}_{||1} =-[2 \epsilon \cos\theta+1.6\epsilon^{3/2}+O(\epsilon^2)] \frac{I{\varphi}'}{B},
\label{nupltot}
\ee
\noindent proportional to ${\varphi}â€™$.   Inserting this into Eq. (\ref{AngMom1}), we get the angular momentum equation in the form
$$ \left(1+2q^2+1.6 \frac{q^2}{\sqrt{\epsilon}} +O(q^2)\right)\partial_t \varphi' =\Tpr  $$
\noindent from which the Rosenbluth-Hinton effective mass is seen to be the factor 
multiplying $\partial_t \varphi'$. The $(1+2q^2)$ factor is the well known Pfirsch-Schluter factor \cite{HK} arising from the circulating particles response. We can see that the $ 1.6\, q^2/\sqrt{\ep}$ is the dominant term.  It can be checked that effective mass contributions are, respectively, $ 1.2\, q^2/\sqrt{\ep}$ due to the TPs and  $ 0.4\, q^2/\sqrt{\ep}$ due to the CPs.   
  
\subsection{Rosenbluth-Hinton $||$ flows }
We note from the above that the effective mass is smaller than what we expect from the toy model, given the rapid TP toroidal precession.  We also note that the parallel flow, (\ref{nupltot}), is much less than the toroidal precession speed expected of the TPs.  In particular, given the large precession, we expect a much larger angular momentum contribution from the TPs (even given the lower density fraction of this species). To examine this further, we calculate separately for trapped and circulating species the parallel flows resulting from the RH solution.  Using Eq (\ref{nupl}) and integrating only over $\cE > \mu B_{max}$ , we get for circulating particles (CP)
\ba
(n{u}_{||})^{CP} =n_0[-2 \epsilon \cos\theta+(-1.6+(1+\cos\theta)^{3/2})\epsilon^{3/2}+O(\epsilon^2)]\frac{I {\varphi}'}{B}
\label{CPx}
\ea
\noindent where we have used an expansion  in $\epsilon$.  This flow speed is as expected.   Correspondingly, for the trapped particles (TP), we integrate inside the separatrix over $\mu B_{min}< \cE < \mu B_{max} $  to get for the parallel flow,


\begin{equation}
(n{u}_{||})^{TP}=n_0\left[ -(1+\cos\theta)^{3/2}\epsilon^{3/2}+O(\epsilon^{5/2})\right]\frac{I {\varphi}'}{B}
\label{TPx}.
\end{equation}

The total parallel flow is obtained by summing these \cite{Xiao}, giving
\be
{u}_{||} =-[2 \epsilon \cos\theta+1.6\epsilon^{3/2}+O(\epsilon^2)] \frac{I{\varphi}'}{B},
\label{TOTx}
\ee
in agreement with Eq (\ref{nupltot}). We would expect to see a large toroidal precession from the TPs. Instead, we find the flow of the TP to be smaller than the toroidal precession drift of the TPs by a factor of $\epsilon$.   Further, if we calculate the poloidal velocity of the trapped particle fraction, we find 
\begin{align}
\bm{ {U}^{TP}}\cdot \hat{\theta} = ({u}^{TP}_{||}+\bm{ {U}_E})\cdot \hat{\theta}=\frac{B_p}{B}\left({u}^{TP}_{||}+\frac{I {\varphi}'}{B}\right) \approx \frac{B_p}{B}\frac{I {\varphi}'}{B_0} \approx U_E
\end{align}
the trapped particles seem to have a nonzero bounce averaged poloidal flow. This is puzzling, since for adiabatic changes we expect TPs to have a purely toroidal flow.

Incidentally, we can use $f_0$ and $f_1$ above to calculate the density.  To lowest order, for $f_0=f_0(\cE)$, the density is constant along the magnetic surface.   To first order, $f_1$ is antisymmetric in $v_{\parallel}$, yielding no change in the density.  Likewise, changes in parallel and perpendicular pressures are also zero. The elements of the electron pressure tensor can be used to a posteriori calculate $E_{||}$, as defined in Eq (49) of the Kulsrud \cite{Kulsrud} manuscript.  
[For massless electrons, $E_{||}$ is given essentially by the generalized adiabatic electron response, viz., $neE_{||} = -  \bm{b \dl : P_e}$.]     We find that $E_{||}=0$ to lowest order and also zero to first order given the $f_1$ symmetry. This self-consistently justifies the neglect of $E_{||}$ in our calculation above.

In order to understand the discrepancy between the RH solution and the expected TP contribution to the flows and effective mass, we will now take a different approach to solve the low frequency RH problem.  

\section{Shifted coordinates}
\label{shift}
\begin{figure}
\centering
\begin{subfigure}{.5\textwidth}
  \centering
  \includegraphics[width=0.7\linewidth]{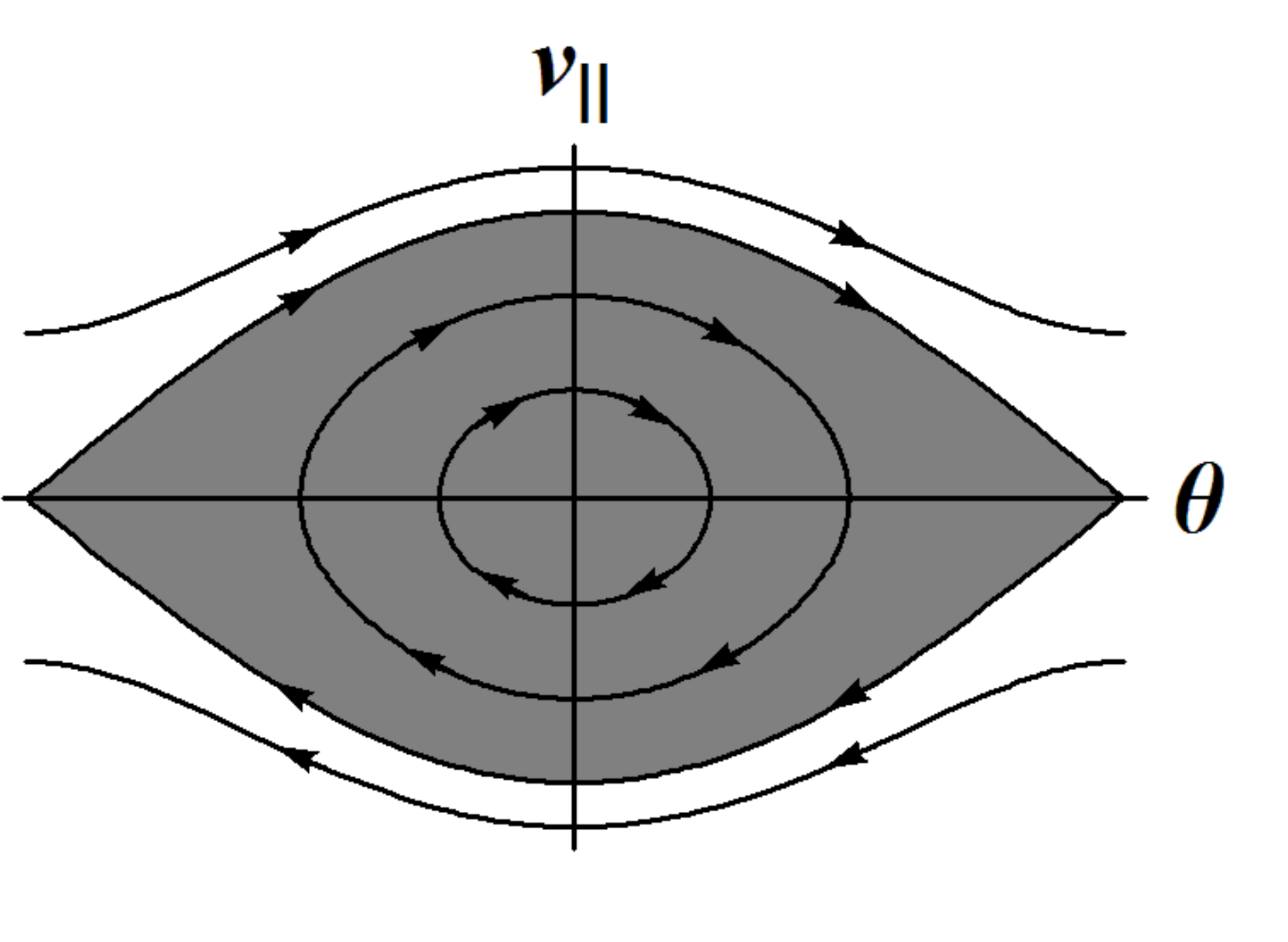}
  \caption{Contours of constant $\cE$ }
  \label{fig:e}
\end{subfigure}%
\begin{subfigure}{.5\textwidth}
  \centering
  \includegraphics[width=0.7\linewidth]{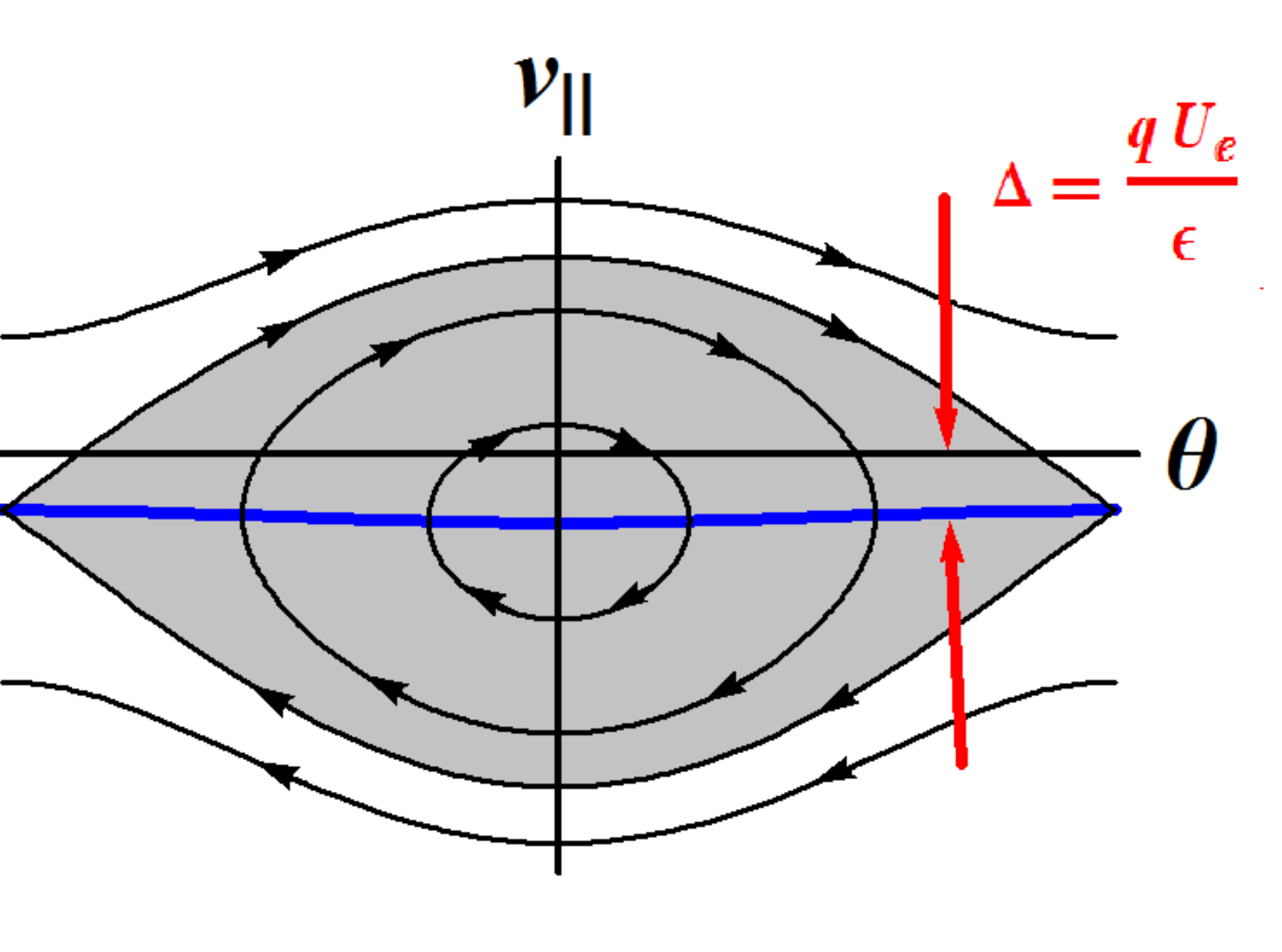}
  \caption{Contours of constant $\cEs$}
  \label{fig:es}
\end{subfigure}
\caption{Comparison of the coordinate systems. [\ref{fig:e}]  standard energy coordinates $(\text{sign}(\vpl),\cE,\theta)$.  [\ref{fig:es}] the shifted coordinates $(\text{sign}(\vpls),\cEs,\theta)$.The shift is $\Iph/B$}
\label{fig:ees}
\end{figure}

As shown earlier, assuming axisymmetry we can rewrite the DKE as
\begin{align}
\frac{\partial f}{\partial t}+ \left(\vpl +\frac{\Iph}{B}\right)\dpl f + 
 \left(v_{||}\frac{\Iph}{B} - \mu B \right)\frac{\dpl B}{B}\frac{\partial f}{\partial v_{||}} =0    
\end{align}
where $f=f(\vpl,\mu,\bm{x},t)$.  We reiterate that this equation is valid for large $q$ and with $U_E$ ordered to be commensurate with $v_{th}/q$.   It can be deduced from this equation, using the method of characteristics, that $$\cEs = \frac{1}{2}v^2_{||}+ \mu B+\Iph \frac{\vpl }{B}$$ is a constant of motion. (See references \cite{kagan,matt,AH} for elaboration on this constant.)  Hence we shift to $(\cEs,\vpls)$ coordinates, defined as follows:
\begin{align}
   \cEs = \frac{1}{2}v^2_{||}+ \mu B+\Iph \frac{\vpl }{B} \quad = \quad \cE +\frac{\Iph \vpl }{B}\nonumber\\ 
\vpls = \vpl + \frac{I \varphi' }{B} \quad \quad \mu \rightarrow \mu, \quad \vec{x} \rightarrow \vec{x}\label{coord} \\
 \dt^3v = \sum_\sigma \frac{B\: \dt\mu\: \dt \cEs}{|\vpls|}\:\,. \nonumber
\end{align}
Here, $\sigma = \frac{\vpls}{|\vpls|}$ denotes the three regions in energy space, namely, the trapped population and the rightward and leftward moving circulating particle populations.   The coordinate $\vpls$ is defined with respect to coordinates shifted downward in $v_{||}$.   $\cEs$ is then a downshifted energy-like coordinate, centered about $\vpls =0$. This shift is depicted in Fig.[\ref{fig:es}].  In $\cEs$ coordinates, centered with respect to $\vpls$,   
the DKE, Eqn(\ref{DKE0}), becomes
\begin{equation}
     \frac{\partial f}{\partial t}+ \vpls \dpl f + \frac{\partial \Iph}{\partial t} \frac{\vpl}{B} \frac{\partial f}{\partial \cEs}=0 
  \label{DKE}
\end{equation}
where $f=f(\cEs,\mu,\psi,\theta,t)$ and ${\partial}/{\partial t}$ is at constant $\cEs$
The angular momentum equation, also recast in $\cEs$ coordinates, is now given by
\begin{align}
\partial_t\left\langle n_0 \varphi' \frac{|\bm{\nabla}\psi|^2}{B^2} \right\rangle  & = \partial_t \left\langle \int \sum_\sigma \frac{B\: \dt\mu \:\dt \cEs}{|\vpls|} \frac{I \vpl}{B} f\right\rangle + \tau_\perp  \:\,. \label{AngMom}
\end{align}
In what follows, we shall use equations (\ref{coord}),(\ref{DKE}),(\ref{AngMom}).

\section{The RH problem revisited }
\label{sec:RHsproblem}
\subsection{Sub-bounce limit }
To make contact with the RH problem, we begin by performing a $\partial_t \ll \omega_b$ expansion, but allowing a large $ U_E \sim \frac{v_{th}}{q}$, which corresponds to a finite downward shift as shown in Fig(\ref{fig:es}). This approach allows for a more transparent calculation.  To dominant order, we have from (\ref{DKE})
\begin{align}
 \vpls  \at{\dpl}{\cEs} f \approx 0 \quad  \Rightarrow   \quad f = f(\cEs,t)
 \label{f_0}.
\end{align}
Annihilating the ${\dpl}_{\cEs}$ operator by bounce averaging gives a constraint equation for $f(\cEs,t)$, viz.
\begin{equation}
 \frac{\partial f}{\partial t}+ \frac{\partial \Iph}{\partial t} \overline{\left(\frac{\vpl}{B}\right)}  \frac{\partial f}{\partial \cEs}=0 \,\: .
 \label{bavg}
\end{equation}
The constraint on $f$ introduces a $\sigma$ dependence. Eqn (\ref{bavg}) and the angular momentum relation (\ref{AngMom}), with $ f = f(\cEs,t)$, form a closed set for the nonlinear $\{f,\varphi'\}$ system.
We now do a subsidiary expansion in small $\frac{\Iph}{B} \ll v_{th}$, denoting $f = f_0 + f_1+ …$ (here, the subscript indices are not the same expansion parameter as in earlier sections).   From (21), the corresponding lowest and first order equations are
\begin{subequations}
\label{subs}
\begin{eqnarray}
\at{\dfrac{\partial f_0}{\partial t}}{\cEs} =0  \quad &\Rightarrow \quad f_0 = f_0 (\cEs) \\ 
\dfrac{\partial f_1}{\partial t}+ \dfrac{\partial \Iph}{\partial t} \overline{\left(\dfrac{\vpl}{B}\right)}  \dfrac{\partial f_0}{\partial \cEs} =0  &\Rightarrow \quad f_1= - \Iph \overline{\left(\dfrac{\vpls}{B}\right)}  \dfrac{\partial f_0}{\partial \cEs}
\end{eqnarray}
\end{subequations}
\noindent where the overbar corresponds to the bounce average holding $\cEs,\mu$ constant, and we note that $\overline{(\vpls / B)}$ = 0 for TPs.   
 
\subsection{RH flows }
We can now calculate the RH flows from Eqns (\ref{subs}). For general $(\Iph/B)/v_{th}$, the dominant order parallel flow for either the TP or the CP populations (or both) is
\begin{align}
 (n U)_{||}=\int \sum_\sigma \frac{B\: \dt\mu\: \dt \cEs}{|\vpls|} \vpl f =\int \sum_\sigma \frac{B\: \dt\mu \:\dt \cEs}{|\vpls|} \left( \sigma |\vpls| -\frac{\Iph}{B}\right) f
 \label{flo}
\end{align}
where, the integrals are to be taken over the appropriate populations and we have used the definition of $\vpls$ as in Eqn \ref{coord} . If we were to expand in small $\varphi'$, correct to first order, we would insert both $f=f_{0} +f_{1}$ in the right hand side integral in (\ref{flo}).  However, since $f_0$ is independent of $\sigma$ for both species, the lowest order term, proportional to $\sigma |\vpls|f_0$, will vanish, by symmetry (with respect to the $\cEs$ coordinates). To first order then, \textit{two} terms must be retained:  one from the $\varphi$ term in the parenthesis and the other from the $f_1$ term.   This yields the expression
\begin{align}
(n U)_{||}=  \int \sum_\sigma \frac{B\: \dt\mu \:\dt \cEs}{|\vpls|} \left(\vpls f_1- \frac{\Iph}{B} f_0\right).
\label{flo2}
\end{align}
\noindent We emphasise that the second term in the integrand appears because of a \lq\lq shift in the Jacobian", and acts on the lowest order $f$.    In particular, even for small $\varphi^\prime$, this term must be retained as it is of the same order as the preceding $f_1$ term. Inserting for $f_1$ in Eqn (\ref{flo2}), we have 
 \begin{align}
(n U)_{||}=- \Iph \int\sum_\sigma B\: \dt\mu \:\dt \cEs \left( \overline{\left(\frac{\vpls}{B}\right)}  \frac{\partial f_0}{\partial \cEs} +\frac{1}{|\vpls| B}f_0\right)
\label{nupls}
 \end{align}
 where, for TPs, we recall that $ \overline{\left(\frac{\vpls}{B}\right)} =0$.   (Eq (\ref{nupls}) can be compared with Eq. (\ref{nupl}), the corresponding equation from the previous section; the latter equation does not have a Jacobian shift term.)
 
Since $ \overline{\left(\frac{\vpls}{B}\right)} =0$ for TPs, the TP parallel flow from Eqn (\ref{nupls}) is 
 
\begin{align}
  (n U)^{TP}_{||}= -\frac{\Iph}{B}\int\limits_{TP} \sum_\sigma \frac{B\: \dt\mu \:\dt \cEs}{|\vpls|}  f_0^{TP}(\cEs,t)= - n^{TP}\frac{\Iph}{B} 
  \label{TPflo}
\end{align}
where $$n^{TP}= n_0\,[\sqrt{\ep(1+\cos{\theta})}+O(\ep^{3/2})]  $$ is the trapped particle density. This parallel flow is a rigid rotor flow and, we note, has an amplitude that corresponds precisely to the precession drift speed.

 We can now also calculate the net poloidal velocity of the TP's,
\begin{align}
\bm{U}^{TP}\cdot \hat{\theta}= \int \dt^3 v\left(\bm{b}\: \vpl +\bm{U_E}\right)\cdot \hat{\theta} ,
\end{align}
 using Eqn (\ref{TPflo}), and we find
\begin{align}
(n\bm{U})^{TP}\cdot \hat{\theta}=\frac{B_p}{B}\left((nU)^{TP}_{||}+ n^{TP}\bm{U_E}\cdot \hat{\theta}\right) =  \frac{\ep}{q}\left((nU)^{TP}_{||}+\frac{\Iph}{B} n^{TP}\right) =0.
\end{align}
This is zero as expected. We note that the ``Jacobian shift" is responsible for resolving the discrepancies.
 
 The CP flow can be calculated from Eqn (\ref{nupls}) assuming that the lowest order distribution function is a Maxwellian. This gives us
 
\begin{align}
(nU)^{CP}_{||} = n_0\frac{\Iph}{B_0} \left[\sqrt{\ep(1+\cos\theta)}-2\ep \cos\theta +O(\ep^{3/2})\right].
\label{CPe}
\end{align}
Note that, to lowest order, the CP flow is a rigid rotor flow,  equal and opposite to the TP flow. Thus from Eqns (\ref{CPe}),(\ref{TPflo}) we see that to dominant order, $(nU)^{CP}_{||}+(nU)^{TP}_{||}\approx 0$.
This says that in the accounting of parallel flows for angular momentum, the large TP precession flow does not materialize as a large parallel flow since it is completely balanced by an oppositely directed CP flow. The $\cos{\theta}$ term in the CP flow is the usual harmonic parallel flow. 

The net poloidal velocity of the CP's is :
\begin{align}
\bm{U}^{CP}\cdot \hat{\theta}= \int \dt^3 v\left(\bm{b}\: \vpl +\bm{U_E}\right)\cdot \hat{\theta}
= \frac{\ep}{q}\left(\frac{\Iph}{B_0} n^{CP} +O(\sqrt{\ep})\right) \approx \bm{U_E}\cdot \hat{\theta}
\end{align}
Hence the poloidal velocity of the CPs is basically the $\bm{E}\times \bm{B}$ flow, consistent with expectations. 
 
 
Summing over the TP and CP flows, we get the total RH flow to be
\be
(nU_{||})^{TP +CP} = -\frac{\Iph}{B_0}\left(2 \:\ep \, \cos \theta + 1.6\, \ep^{3/2} + O(\ep^2)\right)n_0 \:\, .
\label{TOTe} 
\ee
Although the individual flows, Eqns (\ref{TPflo}, \ref{CPe}), differ from the ones obtained using standard neoclassical methods, Eqns(\ref{TPx},\ref{CPx}), the total flows, Eqns(\ref{TOTe},\ref{TOTx}) from our calculation, match the RH solution. Remarkably, the large TP precession flow is balanced by an equally large and oppositely directed flow from the barely CPs.   

\subsection{RH Effective mass}
We now consider the effective mass factor. For this, we would insert $f_0 + f_1$ just found into the 2nd term of the angular momentum equation (\ref{AngMom}).  The second term \newline $\partial_t \left\langle \int \sum_\sigma \frac{B\: \dt\mu \:\dt \cEs}{|\vpls|} \frac{I \vpl}{B} f\right\rangle$  is just the time derivative of the parallel flow, viz., $\partial_t \left\langle \frac{I}{B} (n\,U^{\text{total}}_{||}) \right\rangle $.  A general expression for the parallel flow (for small $\varphi'$) is given by Eq (\ref{nupls}). Inserting this expression as discussed, we obtain the angular momentum equation as
\begin{align}
\partial_t\left\langle \,\varphi' \frac{|\bm{\nabla}\psi|^2}{B^2} \right\rangle  &=  -\partial_t \left\langle \left(\frac{I}{B}\right)^2 \varphi' \left(1 + B \int \limits_{CP} \sum_\sigma \frac{B\: \dt\mu \:\dt \cEs}{|\vpls| n_0} \vpls  \overline{\left(\frac{\vpls}{B}\right)}  \frac{\partial f_0}{\partial \cEs} \right)\right\rangle+\tau_\perp \nonumber
\end{align}
Rearranging, we find
$$\partial_tU_E =\dfrac{F_\perp/m}{1+\cD}$$
where,
\begin{align}
\cD & = \left\langle \left(\frac{1}{B}\right)^2 \left(1 + B \int\limits_{CP} \sum_\sigma \frac{B\: \dt\mu \:\dt \cEs}{|\vpls| n_0} \vpls  \overline{\left(\frac{\vpls}{B}\right)}  \frac{\partial f_0}{\partial \cEs}\right) \right\rangle \left\langle \frac{|\bm{\nabla}\psi|^2}{I^2B^2} \right\rangle^{-1} \label{Deff}\\
& \approx \dfrac{q^2}{\ep^2}\left\langle n\,U^{\text{total}}_{||} \right\rangle/(\Iph/B_0)\nonumber
\end{align}
represents the added effective mass.

To illuminate the role of each species in the effective mass, we  consider the individual effective mass contributions from the TPs and CPs. The TP contribution to the effective mass factor is 

\begin{equation}
\left\langle \left(n\frac{I}{B} U_{||} \right)^{TP} \right\rangle = \partial_t \varphi'\left\langle n^{TP}\left(\frac{I}{B}\right)^2 \right\rangle =n_0[0.9 \sqrt{\ep}+0.15\ep^{3/2}]\partial_t U_E\sim n_0\: \sqrt{\ep}\frac{q^2}{\epsilon^2} \partial_t U_E.
\label{TPEm}
\end{equation} 
Thus the effective mass contribution from the TPs is $\sim O(q^2/\ep^{3/2}) \gg 1 $ as expected from our toy model.
The CP contribution to the effective mass factor is 
\ba
\dfrac{q^2}{\ep^2}\left\langle n\,U^{\text{total}}_{||} \right\rangle/(\Iph/B_0)
\sim -\frac{q^2}{\epsilon^{3/2}} +1.6 \dfrac{q^2}{\sqrt{\ep}} +O(q^2)
\ea
  To lowest order this is equal and opposite to the TP effective mass. Thus the total effective mass is
$1+2 q^2 +1.6 \dfrac{q^2}{\sqrt{\ep}}+O(q^2)$.

\subsection{Flows and effective masses for truncated distributions}

We have seen that the cancellation of the rapid TP precession flow by an oppositely directed flow of barely CPs explains why the effective mass is smaller than that expected from the TPs alone.  But this finding does not unequivocally address whether a distribution function of only TPs would result in the expected large effective mass.  To address this, we consider the distribution function in (\ref{Deff}) to be populated only for $\cEs < \mu B_{max}$.  For this case, the TP contribution to the effective mass can be seen from Eqn (\ref{TPEm}) to be independent of the details of the distribution. The contribution is found to be $\sim (q^2/\ep^2) \, n^{TP}$. Since there are no CPs, $n^{TP}=n$. Therefore, we find the effective mass to be $q^2/\ep^2$, and the accompanying TP flows to be
\begin{align*}
  U^{TP}_{||}= -\frac{\Iph}{B},  \quad  \quad \bm{U}^{TP}\cdot \hat{\theta} =0.
\end{align*}
These findings are completely consistent with our toy model.

\begin{figure}[ht]
\begin{center}
\includegraphics[height=0.4\textwidth,width=0.8\textwidth,angle=0]{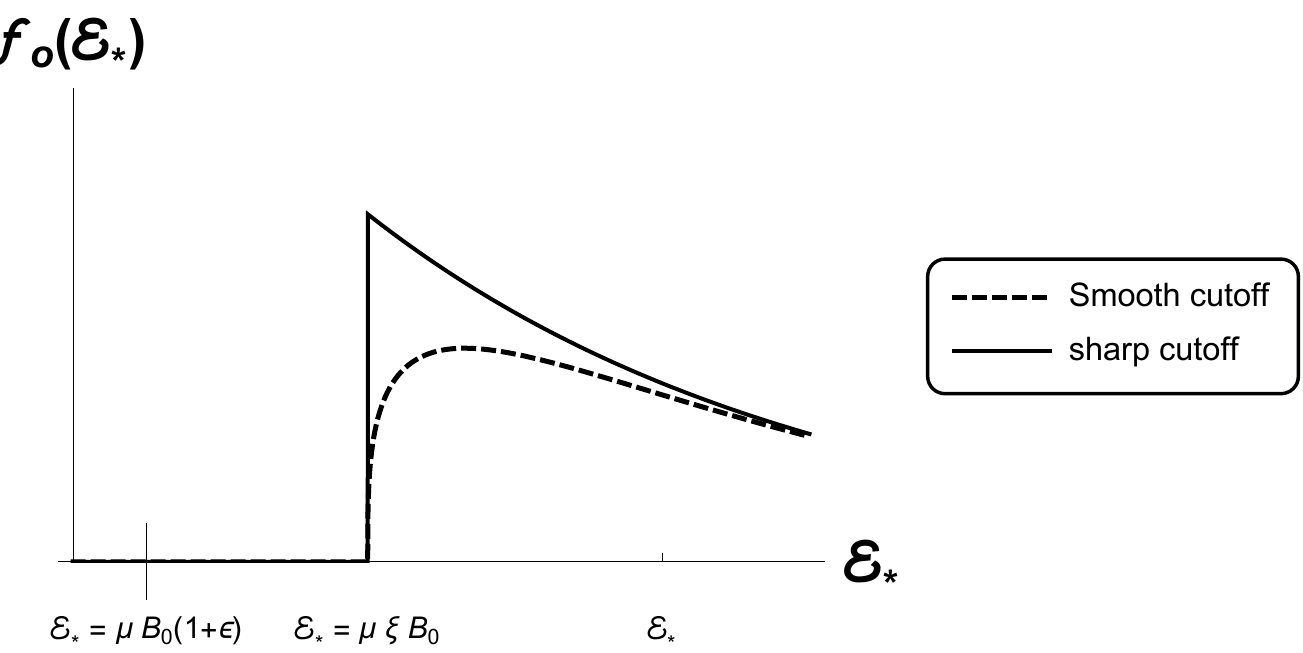}
\caption{The distribution function is nonzero only well above the separatrix $\cEs=\mu B_0(1+\ep)$.}
\label{fig:ecp}
\end{center}
\end{figure}

To complete this line of reasoning, we consider a distribution function with only energetic CPs, i.e., all particles considered to lie well above the separatrix region ($\cEs = \mu B_0(1+\ep)$) in phase space (see solid curve in Fig.[\ref{fig:ecp}] ). The distribution function $f_0(\cEs)$ vanishes both at infinity and at the boundary $\cEs = \mu\: \xi  B_0$, where, $\xi $ is a parameter greater than 1, as shown in Fig. [\ref{fig:ecp}]. To calculate the response to this distribution, we use the following form of Eqn (\ref{nupls}), obtained upon integrating the first term by parts:
\begin{align}
(n U)_{||}= -\Iph \left(\int B\: \dt\mu \:\dt \cEs \left(\frac{1}{|\vpls| B} -\dfrac{\partial}{\partial{\cEs}}\overline{\left(\frac{\vpls}{B}\right)} \right)f_0  - \at{ \int B \dt \mu \overline{\left(\frac{\vpls}{B}\right)} f_0}{\text{CP boundary}} \right).
\label{CPflo}
\end{align}
Using this form, the boundary term in Eqn (\ref{CPflo}) vanishes, and we get
\begin{align}
(n U)_{||}= -\frac{\Iph}{B_0} \int B_0\: \dt\mu \:\dt \cEs \left(\frac{1}{|\vpls|} -B\dfrac{\partial}{\partial{\cEs}}\overline{\left(\frac{\vpls}{B}\right)} \right)f_0  \:\, .
\label{CPflo2}
\end{align}
We evaluate this equation by an expansion in $\ep$.  To lowest order in $\ep$, the RHS is
$$ \dfrac{\partial}{\partial{\cEs}} \overline{\left(\frac{\vpls}{B}\right)} \approx \frac{1}{B_0}\frac{1}{\sqrt{2(\cEs - \mu B_0) }} = \frac{1}{|\vpls|B_0}.$$
Thus, the integral in (\ref{CPflo2}) vanishes to lowest order, indicating that the parallel CP flow is smaller than the TP flow by at least $O(\ep)$.  To evaluate this further, we consider the distribution to be of Maxwellian form but  with a sharp cut-off at $\cEs = \mu \:\xi B_0$ (as shown in Figure \ref{fig:ecp}).    We can then calculate the flows of the ECPs.
\begin{align}
(n U_{||})^{ECP} = -\frac{\Iph}{B_0}\left(  \ep \cos\theta \: h_1(\xi) -(2-\cos 2\theta)\:\ep^2 \:h_2(\xi)\right) 
\end{align}
where $h_1,h_2 $ are simple $O(1)$ algebraic functions of $\xi $.  Further approximating for large $\xi $, we get
\begin{align}
 U_{||}^{ECP} &= -\frac{\Iph}{B_0}\left(  \ep \cos\theta +O(\ep^2)\right)\\
 \bm{U}^{ECP}\cdot \hat{\theta} & \approx \bm{U_E}\cdot \hat{\theta} \: \, .
\end{align}
The effective mass can be shown to be $1+O(q^2)$. These results are consistent with fluid models where we get the oscillating Pfirsch-Schluter flows and the corresponding effective mass factor.  Note that unless we approach the separatrix, there are no $\sqrt{\ep}$ terms. 

\section{ The role of the barely circulating particles}
\label{sec:barelyCP}
\begin{figure}[ht]
\begin{center}
\includegraphics[height=0.4\textwidth,width=0.4\textwidth,angle=0]{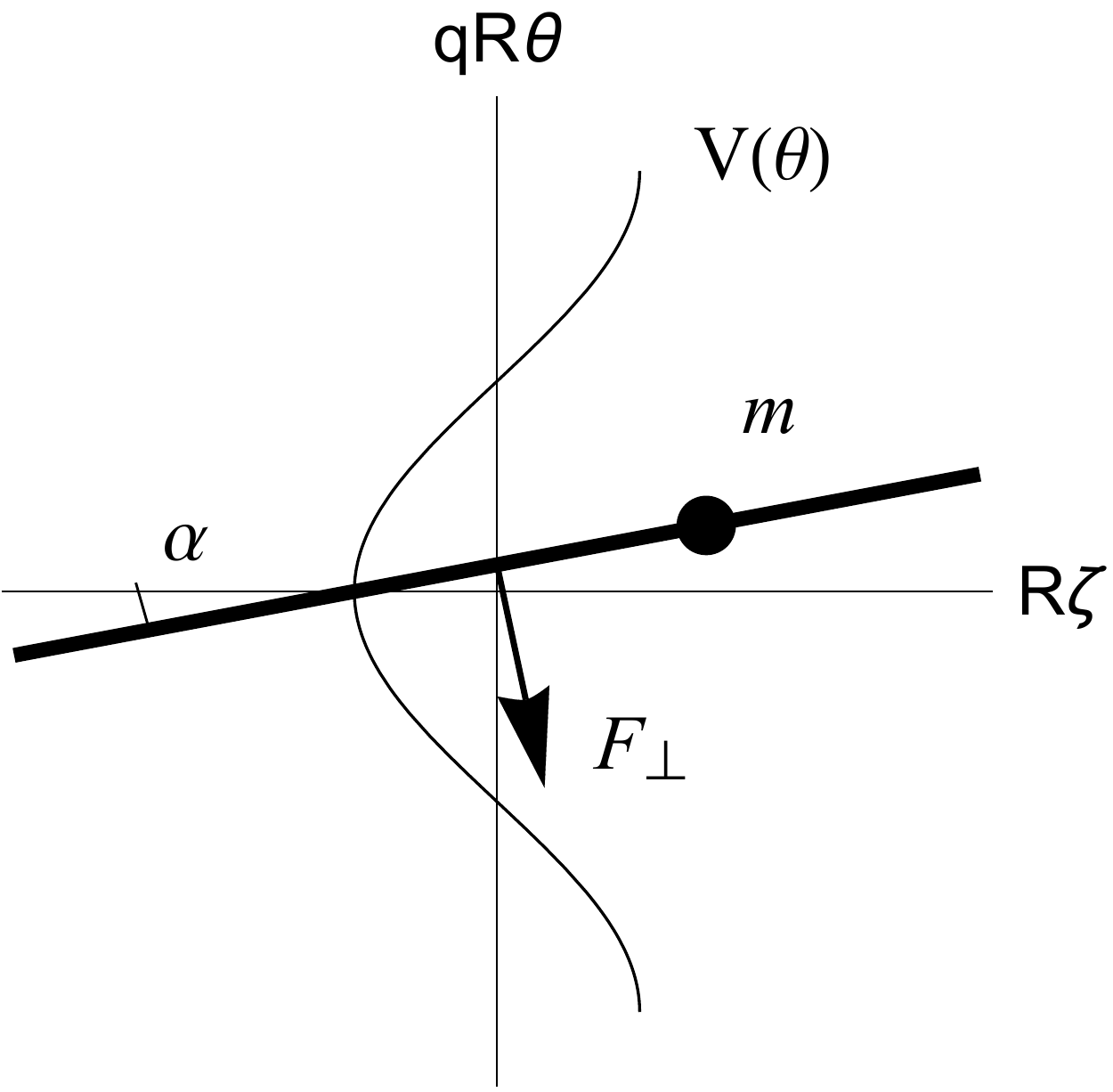}
\caption{New toy model with potential $V(\theta)$}
\label{fig:newtoy}
\end{center}
\end{figure}
We have shown that the large trapped particle precession flow is cancelled to lowest order by an opposite flow from the circulating particles, so that there are no large composite flows of order  $ q U_E/\sqrt{\epsilon}$. We would now like to understand the origin of the opposite flow. We show here that this flow is largely from a class of barely circulating particles.
 To demonstrate this, we begin with a more sophisticated toy model. Consider a particle on a rod as shown in figure[\ref{fig:newtoy}]. The generalized coordinates are $(x=R_0 \zeta, y = q R_0 \theta)$, where $\theta, \zeta$ are analogous to the poloidal and toroidal angles. In addition to being constrained to move only along the rod, the particle also feels a force due to an applied potential $V(\theta)=\mu B(\theta)= \mu B_0 (1-\epsilon \cos \theta )$. Thus, while our previous model allowed only freely circulating particles and deeply trapped particles, our new model allows these but also allows barely circulating particles. 
 
 The Lagrangian is given by, $$ L= \frac{1}{2}\:m R_0^2\left({(q \dot{\theta})^2+(\dot{\zeta}+ q \dot{\theta}\cot{\alpha}})^2\right) -\Blue{\mu m B_0(1-\ep \cos{\theta})} -m R_0^2\: c(t) \: \dot{\zeta} $$
where $c=(\int F_\perp dt )\sin{\alpha}/(m R_0)$ is the impulse due to the applied force $\Fpr\:.$

\noindent The equation of motion is  $$\ddot{\theta}+\omega_b^2 \sin{\theta} = -\frac{\Fpr \cos{\alpha}}{q m R_0} \quad \text{where} \quad \omega_b=\sqrt{\mu B_0 \ep}/qR_0$$
  which shows that our toy model is identical to a driven nonlinear pendulum. We can exploit this similarity to understand the particle trajectories in the presence of the external torque. Let's consider the case where $\Fpr$ is time independent. In this case the work and energy $\cE$ of the driven pendulum is conserved. Thus,
  
  $$\cE = \frac{1}{2}\dot{\theta}^2 - \omega_b^2 \cos \theta + \frac{\Fpr \cos{\alpha}}{q m R_0} \theta = \text{constant.}$$
  
\begin{figure}[ht]
\begin{center}
\includegraphics[height=0.4\textwidth,width=0.4\textwidth,angle=0]{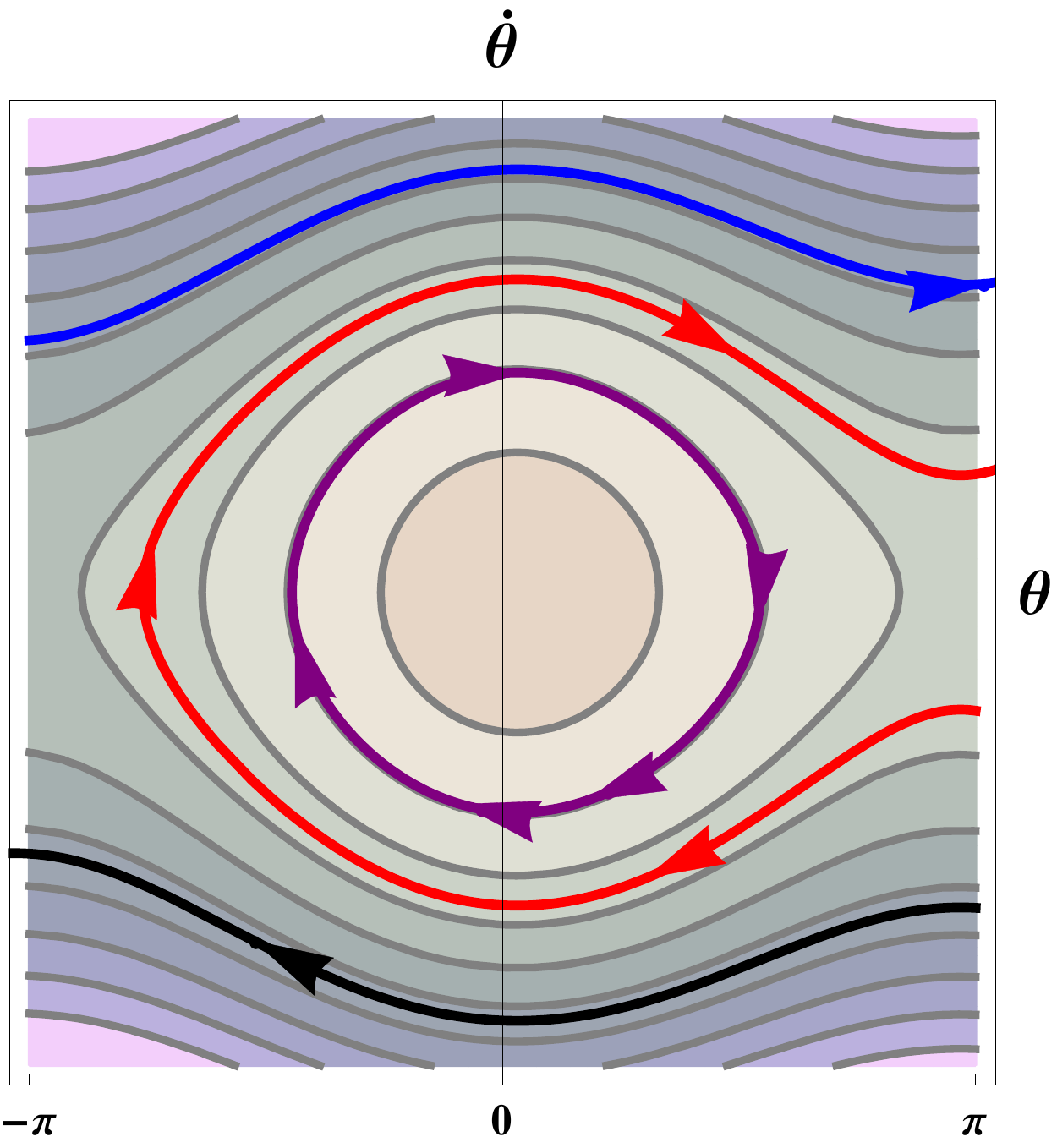}
\caption{driven pendulum phase portrait}
\label{fig:driven}
\end{center}
\end{figure}
\noindent Figure (\ref{fig:driven}) shows the contours of constant $\cE$ for nonzero $\Fpr$. Note that for small $\Fpr$, most of the trajectories resemble the original trajectories of a simple pendulum. However now there exists a group of particles near the separatrix which can change directions. In order to understand the change of direction (sign of $\dot{\theta}$), we note that if we ignore the $\omega_b^2$ term then the equation of motion is simply $\ddot{\theta}= -g$, where $g \propto \Fpr$ is the effective gravity. This means that particles initially moving in the direction of ``g" would not undergo a change in $\dot{\theta}$, but any particle moving opposite to the gravity would slow down and eventually change direction. Figure (\ref{fig:driven}) shows a case where $\Fpr,g<0$ so that eventually $\theta \approx -g t^2$ is positive Thus we can understand how a small applied torque due to $\Fpr$,would generate a flow mostly due to the barely circulating particles. This is actually a very general phenomena. A small perturbation (in this case $\Fpr$) when added to a Hamiltonian system, keeps most of the original trajectories unchanged except for the ones near the separatrix. 
  
  In order to make contact with the drift kinetic system, lets now use the Hamiltonian description of the toy model.  From the Lagrangian we calculate the canonical momenta,
$$\frac{P_\theta}{q m R_0^2}= \dot{\zeta}\:\cot{\alpha}+q \dot{\theta}\;/\sin^2{\alpha}, \quad \quad \frac{P_\zeta}{m R_0^2}= \dot{\zeta}-c +q \dot{\theta}\: \cot{\alpha}.$$
\noindent  Since the Lagrangian is independent of $\zeta$, the ``toroidal angle", $P_\zeta$ must be a constant. The Hamiltonian $\cH$, can now be constructed,
$$\cH/m-\frac{1}{2}\left(\frac{P_\zeta}{m R_0}+R_0 c\right)^2 = \frac{1}{2}\vpls^2+\mu B(\theta)$$
\noindent where, $\vpls \equiv qR_0 \dot{\theta} = (P_\theta-q \cot \alpha (P_\zeta + c\, m R_0^2))/qm$. 
We can now define $\cEs = \frac{1}{2}\vpls^2+\mu B(\theta) $ and write down Liouville's equation for this system in $\{\theta,\zeta,\cEs, P_\zeta \}$ coordinates. We restrict ourselves to the ``axisymmetric" problem by choosing $\partial_\zeta =0$. Thus we have,
\be
\dfrac{\partial f}{\partial t}+\frac{\vpls}{qR_0}\dfrac{\partial f}{\partial \theta}-\dfrac{\partial}{\partial t}( R_0 c\:\cot{\alpha})\:\vpls\:\dfrac{\partial f}{\partial \cEs}=0.
\label{LE}
\ee
 Let us compare the Lioville's equation (\ref{LE}) with the Drift kinetic equation (\ref{DKE}). We note that, by making the identification $\Iph/B  \Longleftrightarrow -R_0\,c\: \cot{\alpha}$, we obtain a one-one relation. This is perhaps not surprising because both equations describe conservation of phase space volume.
 
 Let us now try to understand the large cancellation of the RH flows. From the DKE-toy model equivalence, we see that $\Iph/B \propto - \Fpr$. Figure (\ref{fig:driven}) now corresponds to the case where $\Iph/B>0$. We have already seen that the trapped particles precess with speed $ u_{||}^{TP} \approx q U_E/ \epsilon$. Their net flow from equation (\ref{TPflo}) is $(n u_{||})^{TP} = -n \sqrt{\ep} \Iph/B$  which is negative in this case. The barely circulating particles on the other hand, have a similar density, $\sqrt{\epsilon}$, but have an opposite flow $n \sqrt{\ep} \Iph/B>0$ (see Eqn(\ref{CPe}))
  Thus the two flows cancel. A further explanation of the opposite flows is provided in Appendix(\ref{App:AppendixC}) 
 
\section{Summary}
\label{sec:summary}
 If a tokamak plasma is set into motion with an initial radial electric field $E_r$, the final state, after transients, is a much reduced $E_r$ and a parallel zonal flow consistent with angular momentum conservation.   However, the trapped particle precession angular momentum in the final $E_r$ field is found to be much larger than the zonal flow.   We have shown in this paper that this discrepancy is resolved by the fact that there are reverse flows from the barely passing particles that cancel the large momentum from the TP precession momentum.   Mathematically, we show that, even for small perturbations, there is a linear shift in the Jacobian of the phase space volume element, from $E_r$, that accounts for the reverse flows and the cancellation.   The effective mass for this system is the same as that obtained by Rosenbluth and Hinton \cite{RH} and Xiao et.al \cite{Xiao}. However, the individual contributions from CPs and TPs to the effective mass are very different.

This calculation is done for the completely collisionless response.   As is well-known, the separatrix plays an important role in this problem. In particular, the series expansion in $\ep$ fails near the separatrix because of the logarithmic divergence in the bounce time. Discontinuous flows are obtained. This indicates an inner expansion in the separatrix region to fully understand the RH problem. In a subsequent work we shall use action angle coordinates to address the inner expansion. We shall also show that the RH effective mass can be obtained by simply conserving angular momentum and the second adiabatic invariant. Although, we used MHD ordered drift kinetic equation, the results obtained here can be easily generalized for the drift ordered axisymmetric system. Finally, as is well known \cite{HR}, effects from collisions are also likely to play an important role and act to introduce friction between the large oppositely directed flows.

\section{Acknowledgements}
\label{sec:ackno}
The authors would like to acknowledge stimulating discussion with Prof. T.M.Antonsen Jr, Prof. W. Dorland, Prof. J.F.Drake and Dr. M. Landreman. W.S would also like to acknowledge helpful advice from Prof J.J. Ramos and Prof P.J.Catto. 

This research was funded by the US DOE.

\vspace{0.3in}

\appendix
 \begin{center}
      {\bf APPENDIX}
    \end{center}

\section{Vorticity equation in $\cE$ and $\cEs$ coordinates} \label{App:AppendixA}

We can derive the vorticity equation using $\langle \bm{j}\cdot \grpsi \rangle=0$. From Kulsrud's Eqn\cite{Kulsrud} (46), It can be shown that
\begin{equation}
\partial_t \left\langle n \frac{|\bm{\nabla}\psi|^2}{B^2} \varphi'\right\rangle =\left\langle \int \dt^3v \frac{e}{m}\vec{v_d}\cdot \bm{\nabla}\psi f\right\rangle+\tau_\perp
\end{equation}

In axisymmetric system,
\ba
\vec{B}\times \bm{\nabla}\psi \cdot \bm{\nabla}B= I \vec{B} \cdot \bm{\nabla}B, \quad \at{\dpl}{\cE}\left(\frac{1}{2} \vpl^2\right)= -\mu \dpl B
\ea
and we can show that  $$\frac{e}{m}\vec{v_d}\cdot \bm{\nabla}\psi = \vpl \at{\dpl}{\cE}\left(\frac{I\vpl}{B}\right) = -(\vpl^2+\mu B)I\frac{\dpl B}{B^2} 
$$

Now,
\ba
\cEs= \frac{1}{2}v^2_{||}+ \mu B+\Iph \frac{\vpl }{B} \quad &= \quad \frac{1}{2}\vpls^2+ \mu B- \left(\frac{\Iph }{B}\right)^2\\
\Rightarrow \at{\dpl}{\cEs}\left(\frac{1}{2}  \vpls^2\right) \quad &= \quad -\frac{\dpl B}{B}\left( \mu B + \left(\frac{\Iph }{B}\right)^2\right)
\ea
So,
\ba
\vpls\at{\dpl}{\cEs}\frac{\vpl }{B} &=\vpls\at{\dpl}{\cEs} \left(\frac{\vpls}{B} - \frac{\Iph }{B}\right)\\
&=  -(\vpl^2+\mu B)\frac{\dpl B}{B^2} = \vpl \at{\dpl}{\cE}\left(\frac{\vpl}{B}\right) 
\ea

\section{Proof of equivalence of Angular momentum and vorticity equation}\label{App:AppendixB}
Using the following identities,
\ba
\vec{B}=I\bm{\nabla}\zeta + \bm{\nabla}\zeta \times \bm{\nabla}\psi\\
R^2 \vec{\bm{\nabla}\zeta} \cdot \vec{U_E}= -\frac{|\bm{\nabla}\psi|^2}{B^2}\varphi', \quad R^2 \vec{\bm{\nabla}\zeta} \cdot \hat{b}= \frac{I}{B}
\ea
the angular momentum conservation condition Eqn(\ref{AngMom}), simplifies to
\begin{align*}
\partial_t \left\langle n \frac{|\bm{\nabla}\psi|^2}{B^2} \varphi'\right\rangle -\tau_\perp &= \partial_t \left\langle \int \dt^3v \frac{I \vpl}{B} f\right\rangle\\
&=\left\langle \int \dt^3v \frac{I \vpl}{B}\left(\at{\frac{\partial }{\partial t}}{\cEs}+ \frac{\partial \Iph}{\partial t} \frac{\vpl}{B}  \frac{\partial}{\partial \cEs}\right)f\right\rangle\\
&= \left\langle \int \dt^3v \frac{I \vpl}{B} \left(-\vpls \at{\dpl}{\cEs} f \right)\right\rangle\\
&=\left\langle \int \dt^3v \frac{e}{m}\vec{v_d}\cdot \bm{\nabla}\psi f\right\rangle \quad \text{(By parts)}
\end{align*}

\section{$J_{||}$ invariance and the RH problem}\label{App:AppendixC}
Using method of characteristics we can solve the bounce averaged equations (\ref{bavg})
\begin{align}
\frac{\partial f_0}{\partial t}+ \frac{\partial \Iph}{\partial t} \overline{\left(\frac{\vpl}{B}\right)}  \frac{\partial f_0}{\partial \cEs}=0 \quad \Rightarrow \quad f_0 =f_0\left(J_{||}\right)
\end{align}
 where 
$$J_{||}=\oint \vpl \,\dt l = \oint \vpls \,\dt l - \sigma q R \frac{\Iph}{B_0}$$

is the second adiabatic invariant. Note the crucial sigma dependence \cite{taylor,henrard} for CPs. There is no such sigma dependence in the trapped particle distribution. This means that the CP distribution is not symmetric with respect to $\vpls =0$ and this gives rise to the opposite flows. 
The RH effective mass factor can be obtained directly from the fact that both angular momentum and $J_{||}$ are conserved.


\end{document}